\documentclass[twocolumn]{IEEEtran} 

\usepackage[dvips]{graphicx}

\def\BibTeX{{\rm B\kern-.05em{\sc i\kern-.025em b}\kern-.08em
    T\kern-.1667em\lower.7ex\hbox{E}\kern-.125emX}}

\begin{document}

\title{Exposing Software Defined Radio Functionality\\ To Native Operating System Applications\\ via Virtual Devices} 

\author{Darran Nathan \thanks{The author is with the
DSP Technology Centre, School of Engineering, NgeeAnn Polytechnic, Singapore. (e-mail: darran@projectproteus.org [Darran Nathan]).}}

\markboth{Project Proteus}
{Murray and Balemi: Using the Document Class IEEEtran.cls} 

\maketitle

\begin{abstract}
Many reconfigurable platforms require that applications be written specifically to take advantage of the reconfigurable hardware. In a PC-based environment, this presents an undesirable constraint in that the many already available applications cannot leverage on such hardware. Greatest benefit can only be derived from reconfigurable devices if even native OS applications can transparently utilize reconfigurable devices as they would normal full-fledged hardware devices. This paper presents how Proteus Virtual Devices are used to expose reconfigurable hardware in a transparent manner for use by typical native OS applications.
\end{abstract}


\section{Introduction}
\label{sectIntroduction}

Software Defined Radio (SDR) \cite{bibSDR} refers to the concept of implementing various broadcast / telecommunications standards in software, and running them on the same general purpose hardware. The hardware may be general purpose processors such as the Intel Pentium series, or may be reconfigurable hardware processors such as the Xilinx Virtex series Field Programmable Gate Arrays (FPGA). The promise of SDR has been that devices will become easily adaptable and upgradeable, since only a Software Module download is needed for the device to support a different broadcast / telecommunications standard.

Project Proteus \cite{bibProteus} was initiated by the DSP Technology Centre of NgeeAnn Polytechnic (Singapore) to develop a low-cost PC-based reconfigurable computing platform, one application of which is SDR.

A common limitation of many reconfigurable platforms is that only applications which are fully aware of the existence of this reconfigurable hardware are able to take full advantage of it. For example, an application would normally have to be written to specifically utilize the Application Programming Interface (API) exposed by the Proteus Software Platform (PSP) \cite{bibPSPpaper} to control, reconfigure, or exchange data with the reconfigurable device. This represents a major barrier to being able to realize the full benefit offered by a PC-based reconfigurable computing platform.

Greatest benefit can only be derived if even typical native applications written for the Operating System (OS), and not aware of the underlying PSP, can still utilize this functionality. For example, the Internet Explorer program running on Windows XP should be able to connect to the Internet transparently, using the Proteus Platform with a Modem Software Module downloaded. It can only do so if the OS is able to view the Modem Software Module deployed by the PSP as a full-fledged hardware modem device.

This paper introduces the use of 'Proteus Virtual Devices' to virtualize the existence of hardware, corresponding to the downloaded Software Module on the PSP, to the underlying OS. This allows typical OS applications to transparently utilize the reconfigurable platform as though a corresponding full-fledged hardware device actually exists.

Section \ref{sectPspArch} introduces the architecture of the PSP, Section \ref{sectPvdd} describes how a 'Proteus Virtual Device' exposes reconfigurable hardware functionality to native OS applications, Section \ref{sectExample} illustrates its use with a Virtual Modem Device and the Windows Hyperterminal application, and finally Section \ref{sectConclusion} concludes the paper.

\section{Proteus Software Platform Architecture}
\label{sectPspArch}
\begin{figure}[htb]
\begin{center}
\includegraphics[width=0.2\textwidth]{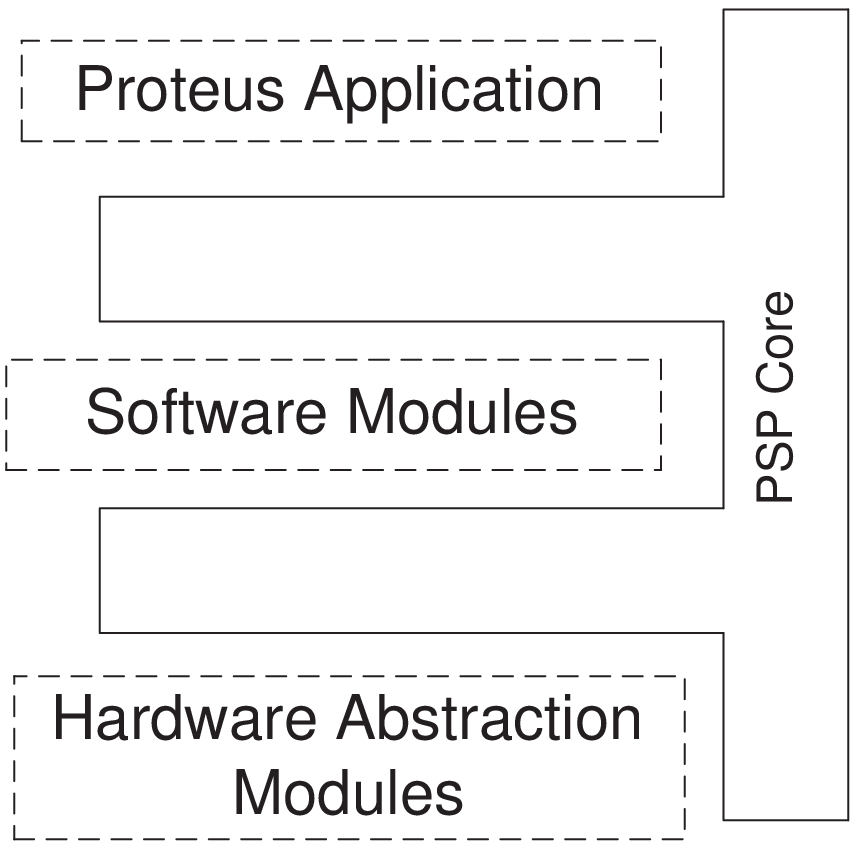}
\caption{Components of the Proteus Software Platform}
\label{figPspInterface}
\end{center}
\end{figure}

The Proteus Software Platform (PSP) has been divided into four main component blocks: the PSP Core, which holds the common set of interfaces and functionality, and three other components: the Proteus Application, Hardware Abstraction Modules (HAMs), and Software Modules. This segmentation is illustrated in Figure \ref{figPspInterface}.

A Hardware Abstraction Module (HAM) serves as the layer of abstraction between the PSP core and the underlying hardware. Any hardware device can therefore be utilized by the PSP through a HAM.

A Software Module is a package of one or more algorithms that are deployed to the reconfigurable hardware by the PSP, as and when desired by the end-user. The Software Module may be a simple algorithm such as a Fast Fourier Transform, or a full telecommunications standard such as GSM or Bluetooth.

When a Software Module has been selected by the user for deployment, the PSP will resolve the type of hardware available (such as FPGAs) via the HAM, and match them with compatible algorithm implementations from the Software Module.

The Proteus Application represents the high-level PC application that has been written to utilize the PSP API to control the platform operation, and to present a graphical user interface to the end-user.

However, we wish to avoid this need for a user to interact with the reconfigurable platform via the user-interface of the Proteus Application. The next section describes how this transparency of use is introduced via the Proteus Virtual Devices.

\section{Proteus Virtual Device Architecture}
\label{sectPvdd}
\begin{figure}[htb]
\begin{center}
\includegraphics[width=0.3\textwidth]{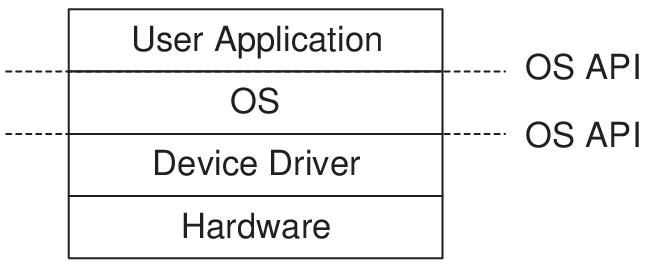}
\caption{Data exchange layers between User Application and Hardware}
\label{figAppLayers}
\end{center}
\end{figure}

An Operating System abstracts underlying hardware from native user applications. These applications indirectly utilize hardware via the API exposed by the OS. At the hardware end, a device driver has to be written to allow the OS to control and exchange data with the hardware. This set of layers and interfaces is illustrated in Figure \ref{figAppLayers}.

As an example, when an email client application wishes to connect to the Internet, it signals this requirement to the OS. The OS then identifies a suitable hardware device that can establish an Internet connection, and utilizes the corresponding device driver to perform this task.

\begin{figure}[htb]
\begin{center}
\includegraphics[width=0.4\textwidth]{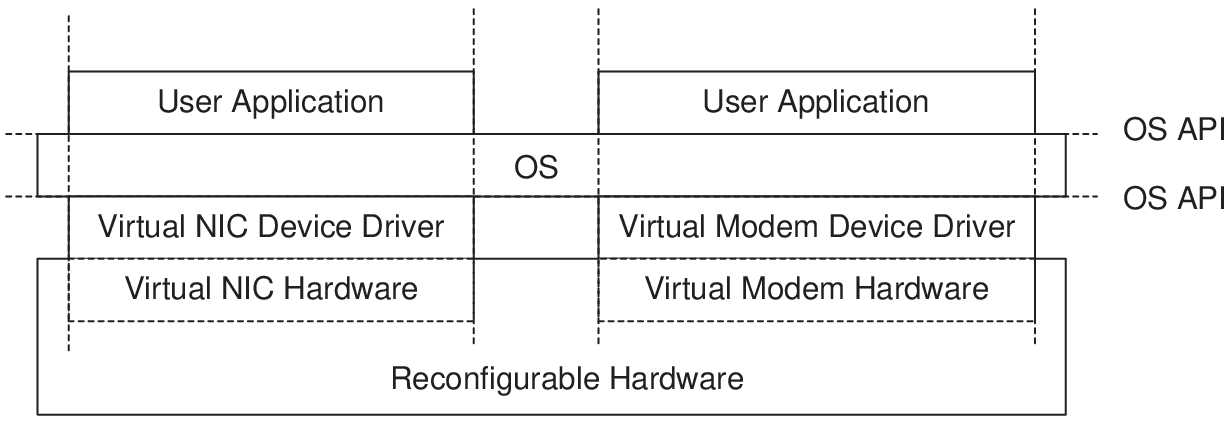}
\caption{Data exchange layers between User Applications and Reconfigurable Hardware}
\label{figReconfAppLayers}
\end{center}
\end{figure}

Therefore, to allow native OS applications transparent use of reconfigurable hardware as full-fledged ordinary hardware devices, a redirection has to be introduced at the 'device driver' layer. This redirection will pass all calls from various device drivers to the same general purpose reconfigurable hardware. Effectively, these device drivers will represent 'virtual' devices because no specific hardware exists for that driver alone - all the 'virtual device' drivers redirect their calls to a common reconfigurable hardware device, which dynamically reconfigures itself according to the requesting driver. This concept is illustrated in Figure \ref{figReconfAppLayers}.

\begin{figure}[htb]
\begin{center}
\includegraphics[width=0.4\textwidth]{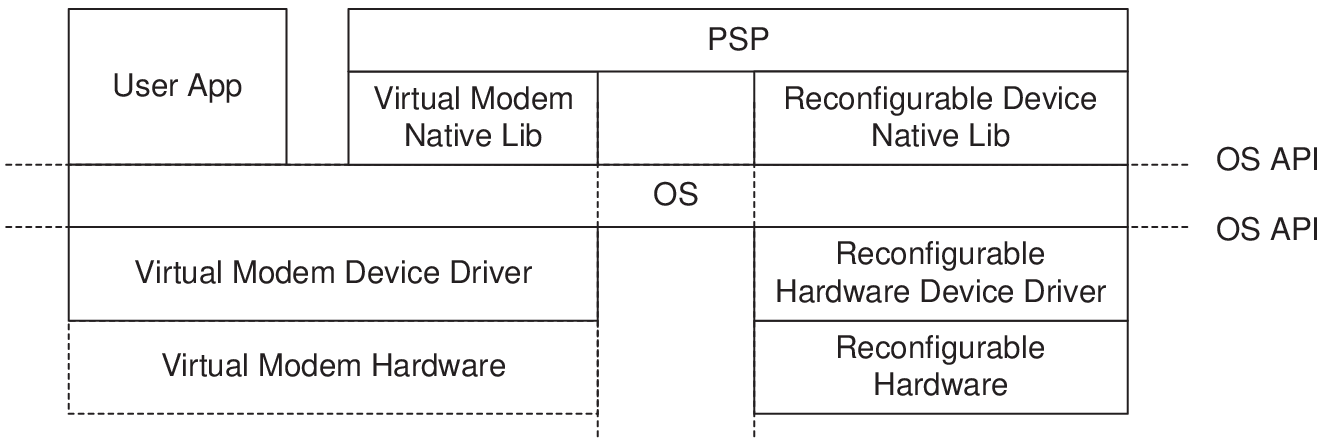}
\caption{Data exchange layers with Reconfigurable Hardware via the PSP}
\label{figPvdAppLayers}
\end{center}
\end{figure}

To arbitrate between multiple requests for use of the reconfigurable hardware, and to control download of the Software Module corresponding to the driver invoked by the OS, all interfacing between the reconfigurable hardware and the virtual device driver is done over the PSP, as shown in Figure \ref{figPvdAppLayers}. The virtual device represented by the OS-specific driver is therefore termed a 'Proteus Virtual Device'.

\begin{figure}[htb]
\begin{center}
\includegraphics[width=0.5\textwidth]{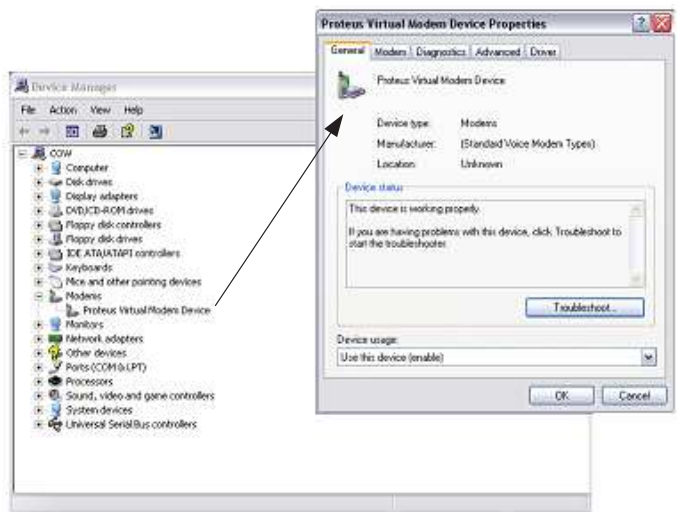}
\caption{Virtual Modem Device in Windows XP}
\label{figDeviceManager}
\end{center}
\end{figure}

\begin{figure}[htb]
\begin{center}
\includegraphics[width=0.4\textwidth]{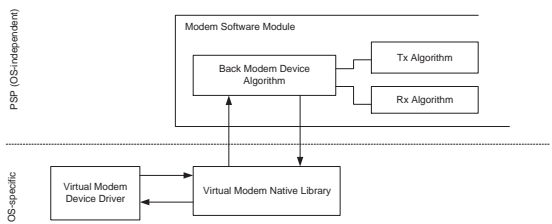}
\caption{Modem Software Module}
\label{figModemSoftwareModule}
\end{center}
\end{figure}

A Proteus Virtual Device therefore appears to the OS to be a full-fledged hardware device, when in fact it redirects all communications to the Software Module downloaded to the PSP. For example, Figure \ref{figDeviceManager} shows the Device Manager of Windows XP with a Proteus Virtual Modem Device driver installed; this appears to the OS as a full-fledged hardware modem device, when internally it actually links up with the Modem Software Module downloaded to the PSP. A simplified illustration of this is shown in Figure \ref{figModemSoftwareModule}.

Each Proteus Virtual Device is distributed as a HAM, which includes the OS-specific driver and an OS-specific native library. The PSP accesses the Proteus Virtual Device driver via the native library.

Since Proteus Virtual Device drivers act as interfaces over which data is exchanged between the OS application and the PSP, there is a need to open two 'handles' to the virtual device, one for the OS application and the other for the PSP.

Each Microsoft Windows Driver Model (WDM) \cite{bibOney} driver will have a 'device object' structure created for it by the OS, to represent the device for which a handle has been opened. A handle can be retained by a single application only, so after a handle has been opened by the native user application, a second 'shadow device object' has to be created for the handle passed to the PSP. The device object associated with the first handle opened by the native user application is called the 'actual device object'. 

\begin{figure}[htb]
\begin{center}
\includegraphics[width=0.5\textwidth]{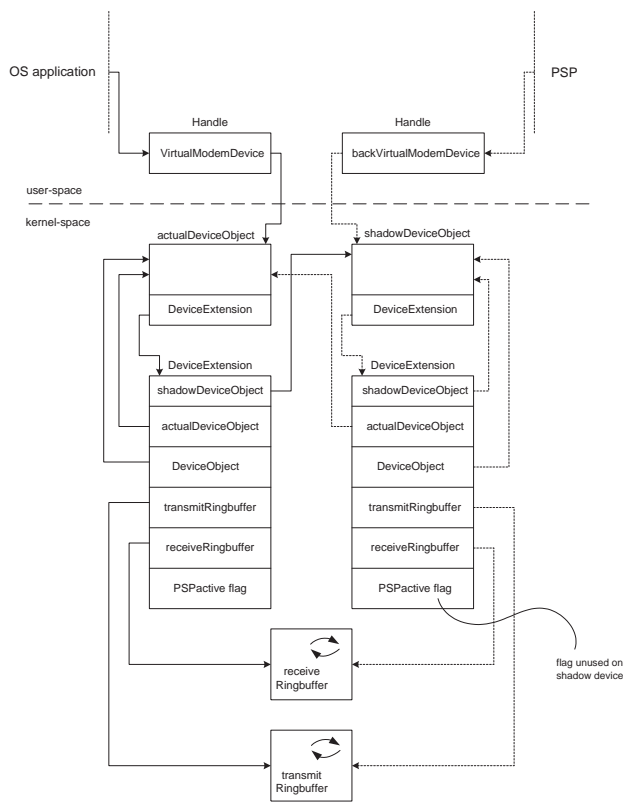}
\caption{Data exchange between an OS application and the PSP, over handles opened to each device object}
\label{figShadowDevice}
\end{center}
\end{figure}

When a read or write operation is performed on a handle, the data is exchanged through ringbuffers shared by both handles / device objects. This is possible because a Windows driver is shared among all device objects created - the same read / write handlers are called, but in the context of the device object corresponding to the handle used. The shadow device object exists for the sole purpose of allowing another handle to be opened - all other fields of the device extension (other than the pointers to the ringbuffers) are not used. Using the Virtual Modem Device as an example, Figure \ref{figShadowDevice} illustrates this set-up of device objects and the data exchange between the native user application and the PSP, via the handles opened to each device object.

\section{Using the Proteus Virtual Modem Device with Microsoft Windows Hyperterminal}
\label{sectExample}
To demonstrate the operation of a Proteus Virtual Device, a Virtual Modem Device HAM has been developed. The Microsoft Windows Hyperterminal application can be used to transparently establish a connection using the Proteus Virtual Modem Device.

The PSP is firstly started to deploy a Modem Software Module. This causes the Virtual Modem Device HAM native library to open a handle to the Proteus Virtual Modem Device, and the OS to create the 'shadow device object'. Once this is done, Hyperterminal can be started.

\begin{figure}[htb]
\begin{center}
\includegraphics[width=0.3\textwidth]{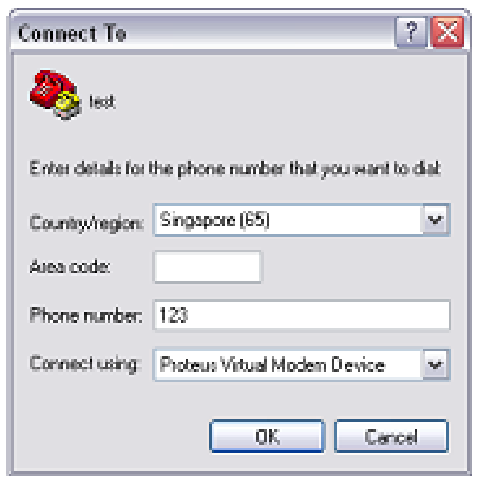}
\caption{Hyperterminal Connect-To dialog}
\label{figHyperterminal}
\end{center}
\end{figure}

Since the OS views the Proteus Virtual Modem Device as a full-fledged hardware device, the 'Proteus Virtual Modem Device' can be selected as the device to be used in the Hyperterminal 'Connect To' dialog, as shown in Figure \ref{figHyperterminal}. Entering any number to dial and clicking 'OK' will start the connection process. This causes the second handle to the device to be opened by Hyperterminal, and the OS to create the 'actual device object'.

\begin{figure}[htb]
\begin{center}
\includegraphics[width=0.5\textwidth]{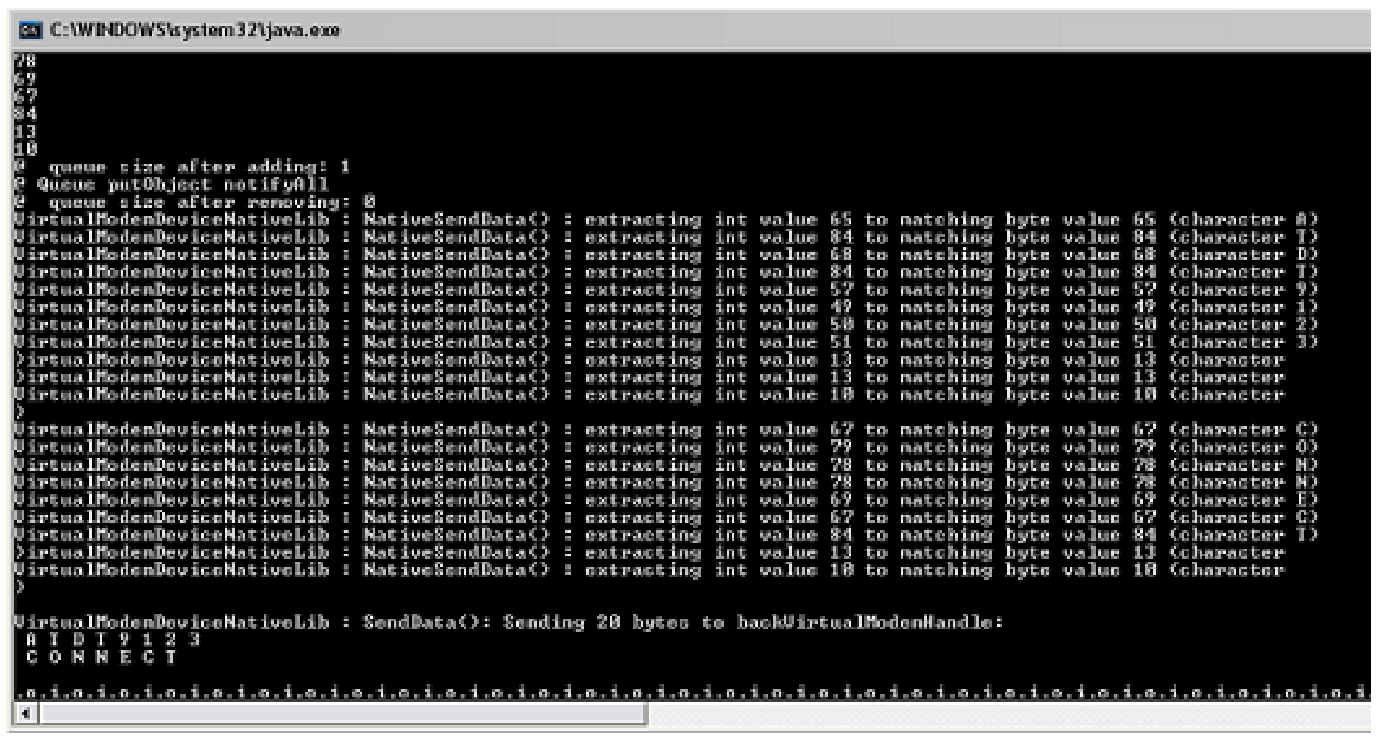}
\caption{PSP debug output during connection establishment}
\label{figPSPdebug}
\end{center}
\end{figure}

Data exchanged with the PSP can be observed in the java debug output window, as shown in Figure \ref{figPSPdebug}. The simple AT parser provided with the Modem Software Module will return the 'CONNECT' string when it receives the 'ATD' command from Hyperterminal during this connect process.

\section{Conclusion}
\label{sectConclusion}
This paper has described a technique of using Proteus Virtual Devices to expose reconfigurable hardware in a transparent manner for use by typical native OS applications. This feature allows PC-based reconfigurable platforms to be fully leveraged on by the many existing OS applications that were developed to utilize only full-fledged hardware devices.

\section*{Acknowledgments}
Special thanks to the Proteus Team, especially to Philip Wong, Kelvin Lim, Kelly Choo, and Andreas Weisensee. Thanks also to Chua Beng Koon and Lim Choo Min of NgeeAnn Polytechnic's Electronic and Computer Engineering Division, for their support of this project. This work was funded by the NgeeAnn Kongsi (Singapore) and NgeeAnn Polytechnic's Innovation \& Enterprise Office. 

\nocite{*}
\bibliographystyle{IEEE}

%



%

\end{document}